\documentclass[12pt]{article}
\usepackage[export]{adjustbox}
\usepackage{epsfig}
\usepackage[english]{babel}
\usepackage{amsfonts,amsmath,mathrsfs,stmaryrd,amsthm}

\usepackage{rotating}
\usepackage{graphicx}
\usepackage{soul,color}
\usepackage{bm}
\usepackage{float}
\usepackage{tabularx}
\usepackage{booktabs}
\usepackage{epsfig}
\usepackage{natbib}
\usepackage[title]{appendix}
\usepackage{multirow} 
\usepackage{verbatim}
\usepackage{hyperref}
\usepackage{latexsym, graphicx, alltt}
\usepackage{graphicx}
\usepackage{setspace}
\usepackage{amssymb}
\usepackage{rotating}
\usepackage{tabularx}
\usepackage{caption}
\usepackage[table]{xcolor}
\usepackage{afterpage}
\usepackage{mathtools}
\usepackage[caption=false]{subfig}
\usepackage{bm}

  \renewcommand{\notin}{\not\in}
\usepackage{flafter}
\newcount\Comments  
\usepackage{color}
\definecolor{darkgreen}{rgb}{0,0.5,0}
\definecolor{purple}{rgb}{1,0,1}
\newcommand{\kibitz}[2]{\ifnum\Comments=1\textcolor{#1}{#2}\fi}

\begin{document}

\title{
Spatio-Temporal Jump Model for Urban Thermal Comfort Monitoring}
\author{  
Federico P. Cortese, Antonio Pievatolo\\
{\small Institute for Applied Mathematics and Information Technologies ``E. Magenes'', Milan}\\ 
\small{
National Research Council of Italy} \\
{\small federico.cortese@mi.imati.cnr.it} 
}

\maketitle

\vspace*{-0.5cm}
\begin{abstract}

Thermal comfort is essential for well-being in urban spaces, especially as cities face increasing heat from urbanization and climate change. Existing thermal comfort models usually overlook temporal dynamics alongside spatial dependencies. We address this problem by introducing a spatio-temporal jump model that clusters data with persistence across both spatial and temporal dimensions. This framework enhances interpretability, minimizes abrupt state changes, and easily handles missing data. We validate our approach through extensive simulations, demonstrating its accuracy in recovering the true underlying partition.
When applied to hourly environmental data gathered from a set of weather stations located across the city of Singapore, our proposal identifies meaningful thermal comfort regimes, demonstrating its effectiveness in dynamic urban settings and suitability for real-world monitoring. The comparison of these regimes with feedback on thermal preference indicates 
the potential of an unsupervised approach to avoid extensive surveys.

 \noindent \vskip5mm \noindent {

}

\end{abstract}


\section{Introduction}

Thermal comfort is defined as ``the condition of mind that expresses satisfaction with the thermal environment''. It reflects the human perception of surrounding thermal conditions, and plays a critical role in the overall well-being within built environments \citep{Fan2023}. 
As urbanization continues to accelerate, cities now accommodate the majority of the global population, consuming around 75\% of the world's energy and thus contributing to significant environmental heat dissipation \citep{nematchoua2021strategies, wei:2019}. 
Coupled with the increasing frequency of extreme weather events, such as heatwaves, this trend raises concerns about the livability of urban areas. Rising temperatures negatively impact human health, leading to conditions such as fatigue, dizziness, and increased heart rate, with severe cases posing life-threatening risks \citep{wang2019tens}. This highlights the need for implementation of advanced tools for urban planning and management strategies to mitigate urban heat and improve thermal comfort in cities \citep{he2021framework}.

Established thermal comfort measures include the Universal Thermal Climate Index (UTCI) \citep{brode2012predicting}, which models human thermoregulation by incorporating variables such as air temperature, wind speed, humidity, and solar radiation, and the Physiological Equivalent Temperature (PET) \citep{hoppe1999physiological}, that simulates heat exchange between the human body and the environment to estimate thermal comfort. 
While UTCI is well-suited for many outdoor scenarios, it faces challenges in highly variable urban environments, where rapid changes in microclimates and environmental conditions can limit its precision \citep{blazejczyk2012comparison}. PET, though useful in various outdoor settings, is less reliable in situations with high humidity or inconsistent clothing insulation, as these factors affect its accuracy \citep{chen2020concepts}.

Recently, machine learning models for predicting thermal comfort have shown promising results \citep{salamone2018integrated,FARD:2022}. 
These models can capture complex relationships between occupants' thermal feedback and environmental variables without requiring explicit knowledge of the physical interactions between factors \citep{mladenovic2016management,kariminia:2016,liu:2020}. This data-driven approach makes machine learning methods particularly suitable for dynamic and heterogeneous urban environments.
The most commonly used algorithms in this field include Support Vector Machines \citep{zhou2020data,alsaleem2020iot}, Artificial Neural Networks \citep{chai2020using,zhou2020data}, along with Tree-Based models and regression methods \citep{luo2020comparing, jung2019heat}.
One limitation of most of these models is that they are supervised learning methods, which depend on labeled data for training. This means the models require pre-existing, accurate thermal comfort feedback to function effectively, making them less suitable in scenarios where such data is unavailable, difficult to gather, or scarce. Furthermore, there is a notable gap in research addressing the dynamic nature of thermal comfort, particularly its temporal evolution, which is critical for understanding how comfort levels change over time in real-world environments.

Spatio-temporal clustering techniques \citep{ansari:2020} offer a valuable alternative by grouping observations based on both spatial and temporal similarity, without the need for labeled data. 
These unsupervised learning methods might capture complex dynamics in thermal comfort by identifying clusters that reflect varying comfort levels over space and time. 
For example, when studying thermal comfort in urban areas, key environmental variables such as air temperature, wind speed, and relative humidity, recorded at fixed weather stations over specified time intervals, can be used to generate meaningful clusters \citep{chen2023constructing, WEI2022, liu:2020}. 

Building on this approach, we propose a new framework to handle complex spatio-temporal data: the \emph{spatio-temporal jump model} (ST-JM). This model extends the capabilities of standard spatio-temporal clustering by introducing persistence constraints that ensure regions close in space and time exhibit similar behaviors. By doing so, it generates smoother predictions that are particularly valuable for 
monitoring and decision making, reducing sensitivity to abrupt shifts in input data and enhancing overall model stability. 
A key strength of our model is its ability to handle mixed-type datasets, including both continuous and categorical variables, while also managing the common challenges of missing data and unequally sampled time-series. These challenges frequently arise due to sensor malfunctions, transmission errors, or equipment failures at weather stations \citep{kasam2014statistical}.

Our proposal is grounded in the broader class of statistical jump models (JM), a promising class of unsupervised learning methods introduced by \citet{bemporad:2018} as an alternative to hidden Markov models (HMMs) \citep{bart:farc:penn:13, zucchini:2017}. JMs transition between simpler submodels, offering greater flexibility for modeling complex dynamics. \citet{bemporad:2018} demonstrated that a standard Gaussian HMM is a special case within this framework. \citet{nystrup:2020} later incorporated $K$-means clustering to define local submodels, while \citet{nystrup:2021} introduced feature selection for temporal clustering.
Statistical jump models have been successfully applied to various time-series analyses, including cryptocurrency markets \citep{cortese:2023}, equity market analysis \citep{cortese2024generalized, aydinhan2024identifying}, risk management \citep{shu2024downside}, investment strategies \citep{shu2024dynamic}, and air quality assessment \citep{cortese2024statistical}. By extending this framework to spatio-temporal contexts, our ST-JM offers a powerful tool for analyzing dynamic systems across both space and time.

In this work, we make two primary contributions. First, we introduce a novel framework for spatio-temporal clustering that efficiently handles missing data, unequally sampled observations, and complex datasets. Simulation studies show that the ST-JM accurately identifies true clusters, particularly in scenarios with limited data or when clusters are difficult to detect.
Second, we apply this framework to address the critical issue of outdoor thermal comfort monitoring in urban areas. 
To the best of our knowledge, we are the first to study thermal comfort using a dynamic, spatio-temporal approach.
Specifically, we track outdoor thermal comfort dynamics in Singapore using hourly data from 14 weather stations across the city. Our results demonstrate that the model effectively captures and describes the spatio-temporal evolution of thermal comfort and that 
has the potential to predict thermal preferences without
considering any feedback or physiological feature.

The paper is organized as follows: Section \ref{sec:2} outlines the model formulation and its key aspects. In Section \ref{sec:3}, we present a simulation study demonstrating that the proposed model provides highly accurate clustering results, even in scenarios with limited data availability. In Section \ref{sec:4}, we apply the model to analyze the spatio-temporal dynamics of outdoor thermal comfort in Singapore over a one-week period, using hourly weather data and additional derived features. Finally, Section \ref{sec:5}, discusses some limitations and further improvements of the study and draws conclusions.

\section{Methodology}
\label{sec:2}
In this section, we first outline the key concepts of statistical JMs, which form the foundation of the proposed approach. We then introduce the ST-JM, highlighting its main strengths. Finally, we describe an efficient estimation algorithm tailored to this model.
 
\subsection{Statistical jump model}

\citet{bemporad:2018} introduced JMs as a framework for modeling complex systems by combining multiple simpler submodels. 
%
Given a time-series $ \pmb{z}_{t,\cdot}=(z_{t,1},\ldots,z_{t,P})^\prime $, $t=1,\ldots,T$, 
a JM is the result of the minimization of a loss function with the following general form
\[
\sum_{t=1}^{T-1} \left[ l_{\text{mode}}(\pmb{z}_{t,\cdot},\pmb{\theta}_{s_t}) + l_{\text{trans}}(\lambda,{s_t, s_{t-1}}) \right] + l_{\text{mode}}(\pmb{z}_{T,\cdot},\pmb{\theta}_{s_T}),
\]
where $l_{\text{mode}}$ and $l_{\text{trans}}$ are terms for local model fit and transition, respectively. The latent state sequence $s_1,\ldots,s_T$ determines which state is active at each time step, and $\pmb{\theta}_{s_t}$ represents state-specific parameters. The hyperparameter $\lambda$ controls the number of state transitions.

\citet{bemporad:2018} showed that this structure generalizes several known models, including Gaussian HMMs. 
\citet{nystrup:2020} focuses on a specific case of the general loss function by modifying the transition term to an indicator function and applying a quadratic loss for feature sets,
relating state classification in HMMs to clustering in high-dimensional spaces. Their proposed loss function is
\[
\sum_{t=1}^{T-1} \left[  \| \pmb{z}_{t,\cdot} - \pmb{\mu}_{s_t} \|_2^2 + \lambda \mathbb{I}({s_t \neq s_{t-1}}) \right] + \| \pmb{z}_{T,\cdot} - \pmb{\mu}_{s_T} \|_2^2,
\]
where 
$\pmb{\mu}_{s_t}$ is the mean vector of state $s_t$. Parameters of the HMM can be derived from these quantities.

\citet{nystrup:2020} demonstrated that their algorithm, which alternates between minimizing quadratic loss and solving a dynamic programming problem similar to the \cite{viterbi:1967} algorithm, is robust to poor initialization and distributional misspecification, while also achieving fast convergence.

\subsection{Spatio-temporal jump model}
We position our approach within the framework of geo-referenced (also known as point-referenced) data clustering \citep{paci:2018,ansari:2020}, as it involves time-series data collected at fixed spatial locations.

We define a set of spatial indices \( m = 1, \ldots, M \), where each index corresponds to a spatial point with coordinates \((x_m, y_m)\). For each spatial location, we consider a set of \( P \) features \( p = 1, \ldots, P \), recorded over time. Additionally, we assume that the set of observation times, denoted as \(\mathcal{T} = \{\tau_1, \ldots, \tau_T\}\), consists of observations recorded at 
possibly
unequal intervals.

We consider data represented as a 3-dimensional array
$\boldsymbol{Z}\in \mathbb{R}^{T\times M\times P}$, essentially a collection of $M$ matrices of dimension $T\times P$, denoted as  
$\boldsymbol{z}_{\cdot,m,\cdot}$.
Each of these matrices has rows \(\boldsymbol{z}_{t,m,\cdot} = \left(z_{t,m,1}, \ldots, z_{t,m,P}\right)^\prime\), which represent feature values for location \(m\) at time \(\tau_t\), and columns \(\boldsymbol{z}_{\cdot,m,p} = \left(z_{1,m,p}, \ldots, z_{T,m,p}\right)\), corresponding to values of feature \(p\) at location \(m\) over times \(\tau_1\) to \(\tau_T\).

%
%

We define $\boldsymbol{S}\in \mathbb{R}^{T\times M}$ to be the matrix of states with elements $s_{t,m}$, i.e. the state at time $\tau_t$ and location $m$. We have that $s_{t,m}$ takes values in $\{1,\ldots,K\}$.
We assume that each of the $K$ states is defined by a vector of conditional prototypes \citep{huang1998extensions}, denoted as $\boldsymbol{\mu}_k\in \mathbb{R}^P$, $k=1,\ldots,K$. The term ``prototype'' refers to a generalized representation of the state-specific parameters, which can be either the mean or mode of the data, depending on the type of variables involved.

We propose to fit a spatio-temporal jump model (ST-JM) with $K$ states by minimizing
%
\begin{equation}
 \label{eq:STJM}
\begin{aligned}
f(\boldsymbol{Z};\boldsymbol{\mu},\boldsymbol{S}) = & \sum_{m=1}^M \Bigg[ \sum_{t=1}^{T} \Bigg[ g(\boldsymbol{z}_{t,m,\cdot}, \boldsymbol{\mu}_{s_{t,m}}) 
     -\gamma  \sum_{k=1}^K \sum_{i \neq m}  
    \mathbb{I}(s_{t,i} = k)e^{-\delta_{im}}
     \Bigg]\\
    &  + \sum_{t \in \mathcal{T}^*} \lambda \frac{\mathbb{I}(s_{t+1,m} \neq s_{t,m})}{\tau_{t+1}-\tau_t} \Bigg],
    \end{aligned}
\end{equation}

with respect to $\boldsymbol{\mu}=\{\boldsymbol{\mu}_1,\ldots,\boldsymbol{\mu}_K\}$ and $\boldsymbol{S}$, with $\mathcal{T}^*=\{\tau_1,\ldots,\tau_{T-1}\}$, and $g(,)$ being the \cite{gower1971general} distance. 
%
$\delta_{im}$ denotes the spatial distance between points $i$ and $m$, specifically the geodesic distance.

The term \(\lambda \mathbb{I}(s_{t+1,m} \neq s_{t,m})/(\tau_{t+1}-\tau_t)\) penalizes temporal transitions at location \(m\), 
and the penalty for jumps decreases as the interval length \(\tau_{t+1}-\tau_t\) increases.

The term 
$\gamma \sum_{k=1}^K\sum_{i\neq m} \mathbb{I}(s_{t,i} = k)e^{-\delta_{im}}$ 
reduces the penalty for state \(k\) in proportion to the number of points \(i\) in state \(k\), weighted by their distance from \(m\).  
This penalty favors assignment of the same state, for each spatial location, within a \textit{small} neighborhood\footnote{To improve computational time and efficiency, one might think of restricting the summation to a predefined neighborhood, like $\sum_{i\in \nu_{m}}\sum_{k=1}^K \mathbb{I}(s_{t,i} = k)e^{-\delta_{im}}$, where $\nu_{m} = \{i \neq m : \delta_{im} < D \}$ and \(D\) is a constant.}  around point \( m \).
The exponential decay \(e^{-\delta_{im}}\) ensures that the contribution of each point \(i\) decreases rapidly as the distance \(\delta_{im}\) increases.
%

%

\subsection{Model estimation}

JMs can be efficiently fitted using a coordinate descent algorithm, which alternates between optimizing model parameters for a fixed state sequence and updating the state sequence based on those parameters \citep{nystrup:2020}. In this paper, we adopt a similar approach, optimizing model parameters across the entire dataset while updating the state sequences individually for each spatial point. To enhance spatial similarity, we favour states that frequently occur in the neighborhood of each point during the update process.
This iterative process continues until the state sequence stabilizes or after 10 iterations. While the global optimum is not guaranteed due to initial state sequence dependence, the algorithm is run multiple times with different initializations to improve the solution, selecting the model with the lowest objective value, essentially the $k$-means ++ method of \cite{arthur2007k}.

We address missing data by iteratively imputing missing values with state-conditional prototypes. This is an adaptation of the $k$-pods method proposed by \cite{chi2016k}, which has been shown to be robust even under conditions where the missingness mechanism is unknown, and external information is unavailable\footnote{For additional approaches to handling incomplete mixed-type data, see \cite{aschenbruck2023imputation}.}.

The estimation process can be summarized as follows.
\begin{itemize}
    \item[(a)] Initialize states matrix $\boldsymbol{S}$. Additionally, initialize missing values with unconditional mean or mode of each feature, i.e. 
    $z^{(miss)}_{{ t},m,p}=\boldsymbol{\bar{z}}^{(obs)}_{\cdot,\cdot, p}$ if feature $p$ is continuous, and $z^{(miss)}_{t,m,p}=\operatorname{Mode}\left(\boldsymbol{z}^{(obs)}_{\cdot,\cdot, p}\right)$ if it is 
    categorical. 
    Superscripts $miss$ and $obs$ denote missing and observed, respectively.
    \item[(b)] Iterate for $j \in 1,\ldots$, until $\boldsymbol{S}^j=\boldsymbol{S}^{j-1}$:
    \begin{itemize}
        \item[(i)] Fit model parameters $\boldsymbol{\mu}=\{\boldsymbol{\mu}_1,\ldots,\boldsymbol{\mu}_K\}$
        $$\boldsymbol{\mu}^j=\operatorname{argmin}_{\boldsymbol{\mu}}\sum_{m=1}^M 
        \sum_{t\in \mathcal{T}} g(\boldsymbol{z}_{t,m,\cdot},\boldsymbol{\mu}_{s_{t,m}^{j-1}})
        .$$

        \item[(ii)]
         Update missing values
         $$\boldsymbol{z}^{(miss)}_{ t,m,p}=\mu^j_{s_{ t}, p}, \,\, t=1,\ldots,T, \, p=1,\ldots,P,$$
         where $\mu^j_{s_{ t}, p}$ is the prototype of feature $p$ in cluster $s_{ t}$. 
       
        \item[(iii)]
        For $m=1,\ldots,M$, fit state sequence
%
\begin{equation}
\label{eq:fitseq}
\begin{array}{rl}
\boldsymbol{s}^{j}_{\cdot,m} = & \operatorname*{argmin}_{\boldsymbol{s}} 
\Big\{
\sum_{t=1}^T 
\left[ g(\boldsymbol{z}_{ t,m,\cdot}, \boldsymbol{\mu}^j_{s_{t,m}}) \right. \\
 &\left. -\gamma\sum_{k=1}^K\sum_{i \neq m}\mathbb{I}(s_{t,i}= k)e^{-\delta_{im}}
 \right] 
\\
 & 
 +\sum_{t \in \mathcal{T}^*} \lambda \mathbb{I}(s_{t+1,m} \neq s_{t,m})\frac{1}{\tau_{t+1}-\tau_t}
\Big\}
    \end{array}
\end{equation}

    \end{itemize}
\end{itemize}

In the previous algorithm, at step (b)-(i), we determine the parameters \(\boldsymbol{\mu}\) by analytically minimizing the loss function for each of the \(K\) states.

In (b)-(iii), the most likely sequence of states, for each spatial point $m$, is found 
by means of the following value function, which is defined recursively in time after assigning the optimal $k$ for every time $t$
\begin{equation*}
\label{eq:V}
\begin{array}{rl}
V^{(m)}(T,s) = & g(\boldsymbol{z}_{T,m,\cdot}, \boldsymbol{\mu}_s) 
- \gamma \sum_{i \neq m} \mathbb{I}(s = s_{T,i})e^{-\delta_{im}}
, \\[10pt]
V^{(m)}(t,s) = & g(\boldsymbol{z}_{t,m,\cdot}, \boldsymbol{\mu}_s) + \operatorname{min}_j \left[ V^{(m)}(t+1,j) + \lambda\mathbb{I}(s \neq j)/(\tau_{t+1}-\tau_t) \right] \\
 & 
 - \gamma \sum_{i \neq m} \mathbb{I}(s=s_{t,i})e^{-\delta_{im}}

 .
\end{array}
\end{equation*}
Then the most likely sequence is given by
\begin{equation*}
\begin{array}{rl}
s_{1,m} = & \operatorname{argmin}_j V^{(m)}(1,j), \\[10pt]
s_{t,m} = & \operatorname{argmin}_j \left[ V^{(m)}(t,j) + \lambda \mathbb{I}(s_{t-1,m} \neq j)/(\tau_{t}-\tau_{t-1})
\right], \quad t=2,\ldots,T.
\end{array}
\end{equation*}

The \texttt{R} \citep{R2024} code for estimating the ST-JM model is available on GitHub at \url{https://github.com/FedericoCortese/ST-JM.git}. The estimation process takes on average 1.1 seconds on an AMD A8-5500 APU with Radeon HD Graphics (3.20 GHz, 16 GB RAM).

\section{Simulation study}
\label{sec:3}
We perform simulation experiments to assess the ability of the method to correctly identify dynamic clusters. Following the approach in \cite{paci:2018}, we generate data at \( M \) sparse locations uniformily distributed in $[0,10]^2$, and \( T \) time points using a dynamic space-time model defined as
\[
\boldsymbol{\xi}_{t} = \beta \boldsymbol{\xi}_{t-1} + \boldsymbol{\eta}_{t}
\]
where \(\boldsymbol{\eta}_{t} \in \mathbb{R}^M\) is a Gaussian process with correlation function \( \Gamma(i,j) = \exp\{-\alpha \| (x_i, y_i), (x_j, y_j) \|\} \). Here, \(\alpha\) determines the spatial correlation decay between locations, and \(\beta\) controls the temporal persistence of the process. We set \(\alpha = 0.01\) and \(\beta = 0.90\) as we observed that these values provided an appropriate balance of spatial and temporal correlation in the generated data.

At each time, we slice the process
realization $\tilde{\boldsymbol{\xi}}_{t}$ with respect to $K$ equidistant levels giving rise
to a spatial partition. In particular, we consider $K=3$.
We determine levels using equidistant empirical quantiles of $\tilde{\boldsymbol{\xi}_{t}}$, obtaining $\tilde{\boldsymbol{S}}$.

We then simulate features from a multivariate Gaussian distribution conditional on latent state $\tilde{\boldsymbol{S}}$ 
with \( K \) latent states, defined as
\begin{equation}
    \label{eq:modform}
    \boldsymbol{z}_{t,m,\cdot} \mid \tilde{s}_{t,m} \sim N_P(\boldsymbol{\mu}_{\tilde{s}_{t,m}},  \boldsymbol{\Sigma}_P),
\end{equation}
where the mean vector \(\boldsymbol{\mu}_{s_{t}} = (\mu_k, \ldots, \mu_k)^\prime\in \mathbb{R}^P\) has state-specific elements \(\mu_1 = \mu\), \(\mu_2 = 0\), and \(\mu_3 = -\mu\). The covariance matrix \(\boldsymbol{\Sigma}_P\) has diagonal elements equal to 1, and off-diagonal elements \(\rho_{ij} = \rho\) for \(i,j = 1, \ldots, P\) with \(i \neq j\).
We consider \(\mu = 0.50\) and \(\rho = 0.20\), representing clusters that are challenging to detect due to minimal separation between the state means and slightly correlated features. 

Half of the features (\( P/2 \)) are converted into categorical variables with 3 levels, \( l = 1, 2, 3 \), where the conditional probabilities are defined as
$$
\mathbb{P}(l = j \mid \tilde{s}_{t,m} = k, z_{t,m,\cdot}) = 
\begin{cases} 
0.80 & \text{if } j = k,  \text{and} \, z_{t,m,\cdot}\in (q^{(1)}_{k}, q^{(2)}_k),\\
0.10 & \text{if } j \neq k, \text{and} \, z_{t,m,\cdot} \notin (q^{(1)}_{k}, q^{(2)}_k),
\end{cases}
$$
for \( j = 1, 2, 3 \), 
where $q^{(1)}_{k}$ and $q^{(2)}_k$ are the 10\% and 90\% quantiles of $N_P(\boldsymbol{\mu}_k,\boldsymbol{\Sigma}_P)$.

We evaluate this setup using \( T = \{10, 50\} \) and \( M = \{10, 50\} \), generating 100 datasets for each combination of \( M \) and \( T \). This values are chosen based on the empirical application of Section \ref{sec:5}. 

For the ST-JM model, we evaluate \(\lambda\) and \(\gamma\) values on a grid from\footnote{In the simulations experiments, \(\lambda\) and \(\gamma\) values above 0.10 tend to stabilize the decoded state sequences and parameter estimates, supporting our focus on this range.} 0 to 0.25, with increments of 0.05 for each simulated dataset. The optimal hyperparameters are selected based on maximizing the balanced accuracy (BAC) between true and estimated state sequences, which is defined as

\begin{equation}
\label{eq:bac}
\mathrm{BAC}=\frac{1}{K} \sum_{k=1}^K \frac{t p_k}{t p_k+f n_k},
\end{equation}
with $tp_k$ and $fn_k$ being the number of true positives and false negatives, respectively, in state $k$. We compare the proposal to the $k$-prototype ($k$-prot) method of \citep{huang1998extensions}, which is essentially a ST-JM with $\lambda=0$ and $\gamma=0$.

\subsubsection*{Temporal gaps}

To evaluate the temporal gap penalty \(\tau_{t+1} - \tau_{t}\) in Equation \eqref{eq:STJM}, we artificially introduce gaps by randomly dropping 20\% of the simulated time-series observations. 
We preliminarily increased the sample size so that the final data includes \( T \) observations \textit{after} artificially removing 20\% of the data.

Results in Table \ref{tab:gaps} demonstrate that the ST-JM consistently outperforms the $k$-prot method across various scenarios.
This suggests that the introduction of the jump and spatial penalties is meaningful and non-arbitrary, as they  enhance clustering accuracy.
In fact, for both \( P=10 \) and \( P=20 \), the ST-JM achieves higher BAC, even in scenarios with smaller $T$ and $M$.

The Monte Carlo standard deviations indicate that ST-JM results are stable and reliable, with equal or lower variance than $k$-prot, especially in high-dimensional scenarios. 

\begin{figure}[b]
    \centering
    \includegraphics[width=.9\linewidth]{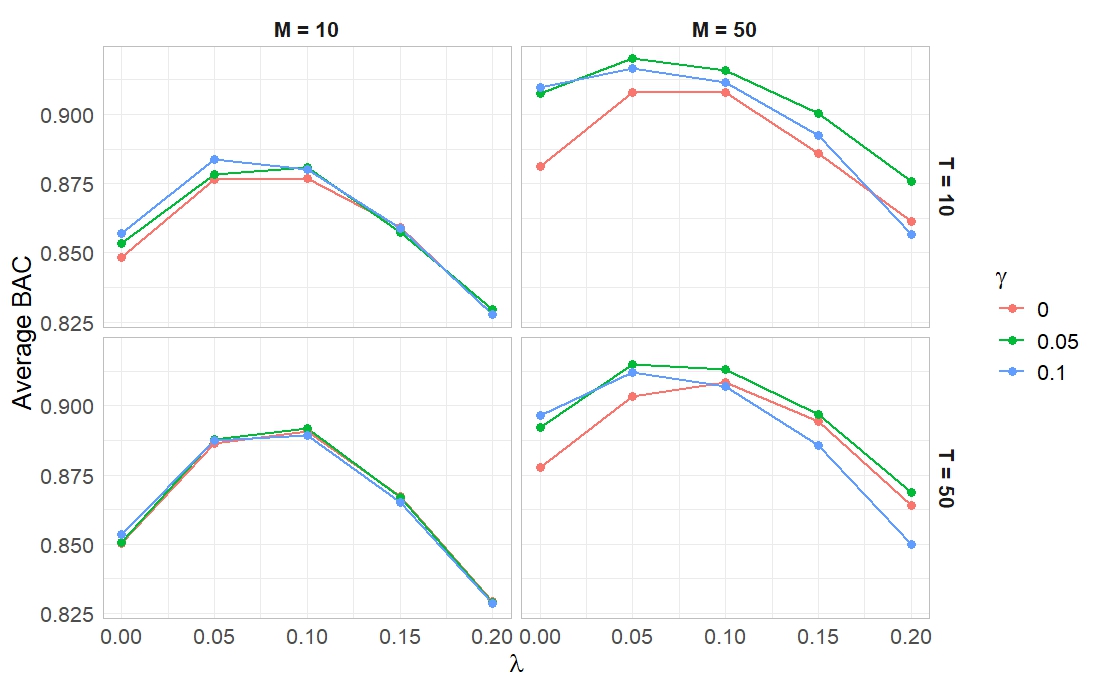}
 \includegraphics[width=.9\linewidth]{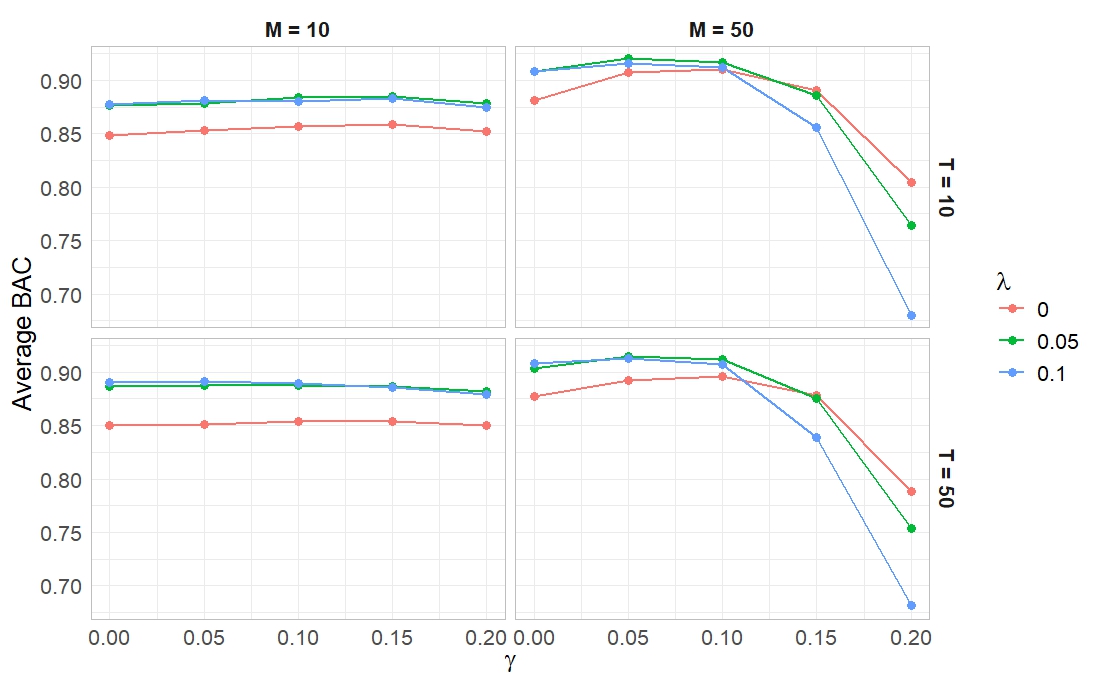}
    \caption{Average Balanced Accuracy (BAC) computed between true and estimated latent sequences, for \(P=10\), and for varying dimensions \( M \) and temporal lengths \( T \).Each subplot presents BAC results across different values of $\lambda$ (top) and $\gamma$ (bottom), with each curve corresponding to a fixed level of $\gamma$ (or $\lambda$, respectively). Missing data were introduced by randomly dropping 20\% of simulated time-series observations, preliminarily adjusting the total number of observations to
preserve dimensionality. }
    \label{fig:simstud_gaps_p10}
\end{figure}

Figure \ref{fig:simstud_gaps_p10} shows average BAC computed between true and estimated sequences when $P=10$, for varying $\lambda$ and $\gamma$. Results show that the optimal values for \(\lambda\) and \(\gamma\) were mostly 0.05, indicating that this configuration works well across most setups\footnote{Additional figures for other simulation scenarios are provided in the supplementary material.}.

\begin{table}[h]
\centering
\caption{
Balanced Accuracy (BAC) computed between true and estimated latent sequences for the different setups: \(P=10\) and \(P=20\). Each subtable shows BAC results for varying dimensions \( M \) and temporal lengths \( T \). Monte Carlo standard deviations are in parentheses. Missing data were introduced by randomly dropping 20\% of simulated time-series observations, preliminarily adjusting the total number of observations to
preserve dimensionality. 
}
\label{tab:gaps}
\begin{tabular}{cccc}
\toprule
\multicolumn{4}{c}{$P=10$} \\
\midrule
$T$ & Method & $M=10$ & $M=50$ \\
\midrule
\multirow{2}{*}{10} & ST-JM   & \textbf{0.89} (0.08) & \textbf{0.92} (0.06) \\
                    & $k$-prot         & 0.85 (0.08)         & 0.88 (0.06)          \\
\multirow{2}{*}{50} & ST-JM   & \textbf{0.89} (0.03) & \textbf{0.92} (0.03) \\
                    & $k$-prot         & 0.85 (0.03)         & 0.88 (0.03)          \\
\midrule
\multicolumn{4}{c}{$P=20$} \\
\midrule
$T$ & Method & $M=10$ & $M=50$ \\
\midrule
\multirow{2}{*}{10} & ST-JM   & \textbf{0.96} (0.05) & \textbf{0.96} (0.05) \\
                    & $k$-prot         & 0.95 (0.05)         & 0.94 (0.06)          \\
\multirow{2}{*}{50} & ST-JM   & \textbf{0.95} (0.02) & \textbf{0.96} (0.02) \\
                    & $k$-prot         & 0.94 (0.06)         & 0.94 (0.03)          \\
\bottomrule
\end{tabular}
\end{table}

\subsubsection*{Missing data}

We assess the accuracy of the ST-JM under missing data conditions by repeating the above procedure, but this time randomly introducing missing values for 5\% and 20\% of the simulated observations.

Results in Tables \ref{tab:5percentMissing} and \ref{tab:20percentMissing} show that the ST-JM consistently outperforms the $k$-prot method. 
For 5\% missing data scenario, the ST-JM maintains high BAC, particularly as 
$M$ and $T$ increase. For instance, with \(M=50\) and \(T=10\), the ST-JM achieves a BAC of 0.89, outperforming $k$-prot result of 0.85. This pattern holds consistently, with ST-JM showing equal or smaller standard deviations, indicating greater stability across different setups. 

With 20\% missing data, although the overall accuracy drops due to the increased missingness, the ST-JM still outperforms $k$-prot. For example, with \(M=50\) and \(T=50\), the ST-JM achieves a BAC of 0.74, compared to $k$-prot’s 0.66. The performance gap between the two methods widens as missing data increases, demonstrating robustness of ST-JM in handling complex scenarios.

As in the previous cases, the optimal values for both \(\lambda\) and \(\gamma\) are 0.05 across most scenarios.

\begin{table}[h]
\centering
\caption{
Balanced Accuracy (BAC) computed between true and estimated latent sequences under 5\% missing data conditions for different setups: \(P=10\) and \(P=20\). Each subtable shows BAC results for varying dimensions \( M \) and temporal lengths \( T \). Monte Carlo standard deviations are in parentheses.
}
\label{tab:5percentMissing}
\begin{tabular}{cccc}
\toprule
\multicolumn{4}{c}{$P=10$} \\
\midrule
$T$ & Method & $M=10$ & $M=50$ \\
\midrule
\multirow{2}{*}{10} & ST-JM   & \textbf{0.84} (0.09) & \textbf{0.89} (0.07) \\
                    & $k$-prot & 0.80 (0.08)         & 0.85 (0.07)          \\
\multirow{2}{*}{50} & ST-JM   & \textbf{0.84} (0.04) & \textbf{0.89} (0.03) \\
                    & $k$-prot & 0.80 (0.04)         & 0.85 (0.03)          \\
\midrule
\multicolumn{4}{c}{$P=20$} \\
\midrule
$T$ & Method & $M=10$ & $M=50$ \\
\midrule
\multirow{2}{*}{10} & ST-JM   & \textbf{0.90} (0.07) & \textbf{0.95} (0.04) \\
                    & $k$-prot & 0.89 (0.07)         & 0.93 (0.05)          \\
\multirow{2}{*}{50} & ST-JM   & \textbf{0.90} (0.03) & \textbf{0.93} (0.03) \\
                    & $k$-prot & 0.88 (0.04)         & 0.91 (0.03)          \\
\bottomrule
\end{tabular}
\end{table}

\begin{table}[h]
\centering
\caption{
Balanced Accuracy (BAC) computed between true and estimated latent sequences under 20\% missing data conditions for different setups: \(P=10\) and \(P=20\). Each subtable shows BAC results for varying dimensions \( M \) and temporal lengths \( T \). Monte Carlo standard deviations are in parentheses.
}
\label{tab:20percentMissing}
\begin{tabular}{cccc}
\toprule
\multicolumn{4}{c}{$P=10$} \\
\midrule
$T$ & Method & $M=10$ & $M=50$ \\
\midrule
\multirow{2}{*}{10} & ST-JM   & \textbf{0.67} (0.10) & \textbf{0.77} (0.08) \\
                    & $k$-prot & 0.63 (0.08)         & 0.73 (0.08)          \\
\multirow{2}{*}{50} & ST-JM   & \textbf{0.68} (0.05) & \textbf{0.74} (0.12) \\
                    & $k$-prot & 0.64 (0.05)         & 0.66 (0.09)          \\
\midrule
\multicolumn{4}{c}{$P=20$} \\
\midrule
$T$ & Method & $M=10$ & $M=50$ \\
\midrule
\multirow{2}{*}{10} & ST-JM   & \textbf{0.75} (0.10) & \textbf{0.84} (0.08) \\
                    & $k$-prot & 0.70 (0.10)         & 0.82 (0.07)          \\
\multirow{2}{*}{50} & ST-JM   & \textbf{0.75} (0.04) & \textbf{0.85} (0.03) \\
                    & $k$-prot & 0.71 (0.04)         & 0.83 (0.04)          \\
\bottomrule
\end{tabular}
\end{table}

\clearpage
\section{Empirical study}
\label{sec:4}

\begin{figure}
    \centering
    \includegraphics[width=0.8\linewidth]{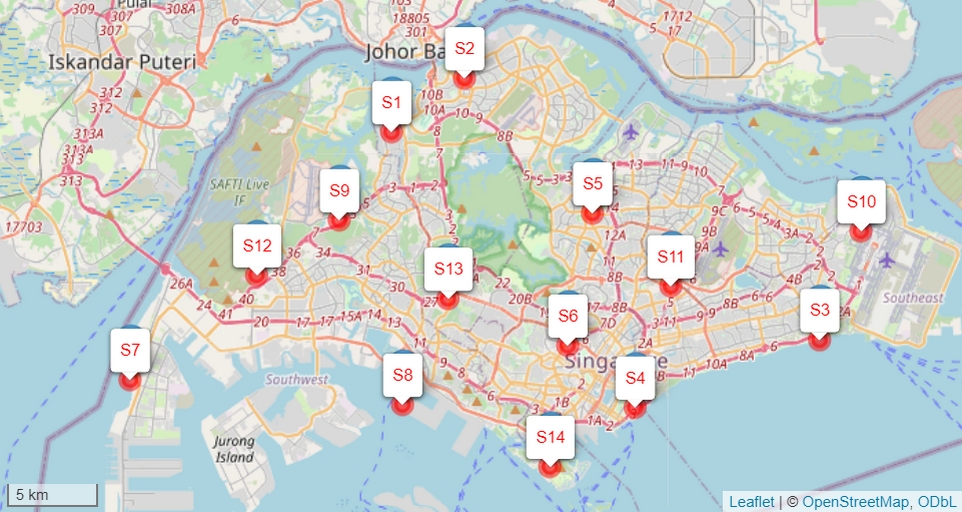}
    \caption{Geographic locations of the 14 weather stations (S1–S14) used for monitoring thermal comfort across Singapore.}
    \label{fig:sing_map}
\end{figure}

In this Section, we study the spatio-temporal evolution of outdoor thermal comfort in the city of Singapore. 
Data is sourced from the Kaggle competition ``Cozie Quiet City''\footnote{This competition includes the collection of 9,808 smartwatch micro-surveys and 2,659,764 physiological and environmental measurements from 98 participants, using the Cozie-Apple platform. These data are combined with urban metrics from the Urbanity Python package \citep{yap2023urbanity,yap2023global, yap2023incorporating}, driven by geolocation and urban digital twin technologies. More information are available here: \url{https://www.kaggle.com/competitions/cool-quiet-city-competition/overview}}. 
In particular, we consider hourly data on air temperature (Temp, °C), relative humidity (RH, \%), rainfall (RF, mm), and wind speed (WS, m/s). Data spans from April 18, 2023, at 12:00, to April 26, 2023, at 7:00, and is recorded at 14 weather station, as mapped in Figure \ref{fig:sing_map}.
During the same period, we also gathered user feedbacks from various locations and times:
this feedback is used solely to evaluate the predictive power of our model, and details on this comparison are given in Section \ref{sec:5}.
We also include UTCI (°C) data obtained from the ``thermal comfort indices derived from ERA5 reanalysis'' database \citep{dinapoli2020thermal}\footnote{This dataset offers a historical reconstruction of human thermal stress and discomfort indices in outdoor conditions, using ERA5 reanalysis data from the European Centre for Medium-Range Weather Forecasts (ECMWF). ERA5 integrates global model data with observations to provide a comprehensive and consistent description of the Earth's climate over recent decades and serves as a reliable proxy for observed atmospheric conditions. The dataset is produced by ECMWF. More information are available at :\url{https://cds.climate.copernicus.eu/datasets/derived-utci-historical?tab=overview}}. Unfortunately, this data is available only as a time-series average for the entire city of Singapore, without spatial resolution.

We also include two categorical features: \textit{windy}, with levels from 1 to 6, for increasing levels of intensity of the wind, determined from the Beaufort scale \citep{saucier1955principles}, and \textit{hour}, the hour of the day.

Additionally, similar to \cite{cortese:2023}, we consider 5-hours moving averages and standard deviations of Temp, RH, RF, WS.  These features smooth short-term fluctuations and highlight mid-term trends, aiding in the analysis of thermal comfort over time \citep{aparicio2023analysis}. 

The final dataset includes \( T = 178 \) observations collected from \( M = 14 \) weather stations, encompassing \( P = 16 \) variables. Regarding time sampling, 4.49\% of the observations are delayed by two hours compared to the previous one, and one observation (0.56\%) is delayed by three hours.

Tropical climate of Singapore is characterized by consistently high temperatures and humidity throughout the year, as reflected in the summary statistics of the main weather variables presented in Table \ref{tab:weather_summary}.
Notably, Temp has a mean value of 28.81°C, with relatively low variability (SD = 1.90°C). 
RH displays a wider range, from 44.32\% to 99.59\%, with an SD of 10.31\%, indicating considerable fluctuations in humidity levels. 
RF shows a mean close to zero, reflecting the low daily average, with occasional peaks (Max = 3.25 mm). 
WS has a higher variability (SD = 2.19 m/s), with some instances of calm conditions (Min = 0.50 m/s) and occasional gusts (Max = 14.92 m/s). 
The UTCI shows moderate variability (SD = 2.75°C), with a range from 25.95°C to 35.93°C. 
Missing values are uniformly distributed across all variables, except for UTCI, which has complete data. 

\begin{table}[h]
\centering
\caption{Mean, maximum, minimum, standard deviation (SD) and percentage of missing values (NA) of the main variables.}
\label{tab:weather_summary}
\begin{tabular}{lccccc}
\toprule
 & Mean & Min & Max & SD & NA (\%) \\
\midrule
Temp & 28.81 & 22.33 & 35.40 & 1.90  &2.81 \\
RH & 79.06 & 44.32 & 99.59 & 10.31  &2.81 \\
RF & 0.02 & 0.00 & 3.25 & 0.18  &2.81 \\
WS & 3.69 & 0.50 & 14.92 & 2.19 &2.81 \\
UTCI &30.62  & 25.95 &35.93   & 2.75&0\\
\bottomrule
\end{tabular}
\end{table}

Table \ref{tab:correlation_matrix} presents the correlation and partial correlation coefficients between the main weather variables. As expected, Temp is negatively correlated with RH (-0.84) and RF (-0.20). WS shows a weak positive correlation with temperature (0.18) and a stronger correlation with UTCI (0.64), indicating that WS might have an impact on perceived thermal comfort. Partial correlations, indicated in italics, provide additional insights, particularly between temperature and UTCI (0.28), showing that temperature remains a significant predictor of thermal comfort, even when accounting for the other variables.

\begin{table}[h]
\centering
\caption{Correlations and partial correlations (in italics) of the main variables.}
\label{tab:correlation_matrix}
\begin{tabular}{lccccc}
\toprule
 & Temp & RH & RF & WS & UTCI \\
\midrule
Temp & 1.00 & -0.84 & -0.20 & 0.18 & 0.64 \\
\textit{RH} & \textit{-0.73} & 1.00 & 0.15 & -0.16 & -0.62 \\
\textit{RF} & \textit{-0.19} & \textit{-0.01} & 1.00 & 0.13 & -0.02 \\
\textit{WS} & \textit{0.08} & \textit{0.01} & \textit{0.16} & 1.00 & 0.20 \\
\textit{UTCI} & \textit{0.28} & \textit{-0.20} & \textit{0.13} & \textit{0.09} & 1.00 \\
\bottomrule
\end{tabular}
\end{table}

Table \ref{tab:station_weather} summarizes the key weather variables measured across 14 stations, revealing some variation in Temp, RH, RF, and WS. Temp remains consistent across the stations, with values around 28-29°C. RH, however, shows more variability, ranging from 73.8\% to 89\%, which could be influenced by local microclimates. RF is generally low, except for a few stations like S9, where it peaks at 0.06 mm. WS also differs notably between locations, with some stations recording higher values, suggesting more exposed areas or wind channels.

\begin{table}[h]
\centering
\caption{Average air temperature (Temp, °C), relative humidity (RH, \%), rainfall (RF, mm), and wind speed (WS, m/s) recorded at each weather station.}
\label{tab:station_weather}
\begin{tabular}{lcccc}
\toprule
Station & Temp (°C) & RH (\%) & Rainfall (mm) & Wind Speed (m/s) \\
\midrule
S1  & 28.5 & 80.0  & 0.03 & 1.82 \\
S2  & 28.4 & 77.0  & 0.02 & 3.57 \\
S3  & 29.3 & 78.7  & 0.01 & 5.11 \\
S4  & 29.5 & 89.0  & 0.02 & 2.45 \\
S5  & 29.0 & 73.8  & 0.01 & 2.82 \\
S6  & 28.6 & 76.8  & 0.01 & 3.87 \\
S7  & 28.8 & 78.6  & 0.05 & 3.10 \\
S8  & 29.1 & 79.9  & 0.01 & 6.97 \\
S9  & 28.3 & 85.0  & 0.06 & 6.01 \\
S10 & 29.2 & 75.2  & 0.00 & 4.89 \\
S11 & 29.3 & 75.2  & 0.00 & 2.96 \\
S12 & 28.1 & 80.3  & 0.07 & 3.26 \\
S13 & 28.6 & 80.4  & 0.02 & 2.10 \\
S14 & 28.9 & 76.9  & 0.01 & 2.70 \\
\bottomrule
\end{tabular}
\end{table}

\subsection{Hyperparameters selection}
The ST-JM model involves three key hyperparameters: temporal persistence \(\lambda\), spatial persistence \(\gamma\), and number of states \(K\), all of which require a selection criterion. In sparse \(k\)-means clustering, \citet{witten2010framework} propose using GAP statistics as a selection method. Meanwhile, in the statistical JM literature, \citet{cortese2023b} and \citet{cortese2024generalized} implement a generalized information criterion for model selection, which has proven effective for time-series data. However, adapting this criterion for spatio-temporal data is not straightforward.

Given this challenge, we adopt a heuristic approach, setting \(K=3\), as this aligns with the number of response categories provided by users during the study period. 
The hyperparameters \(\gamma\) and \(\lambda\) were both fixed at 0.05, a value that was consistently optimal across various scenarios during our simulation studies. 

\subsection{Results}

We characterize each state using state-conditional prototypes of the key features, as shown in Table \ref{tab:state_conditional_prot}. Based on these prototypes, we identify State 1 as a \textit{cool} regime, State 2 as a \textit{neutral} regime, and State 3 as a \textit{hot} regime.

\begin{figure}[b]
    \centering
    \includegraphics[width=.8\linewidth]{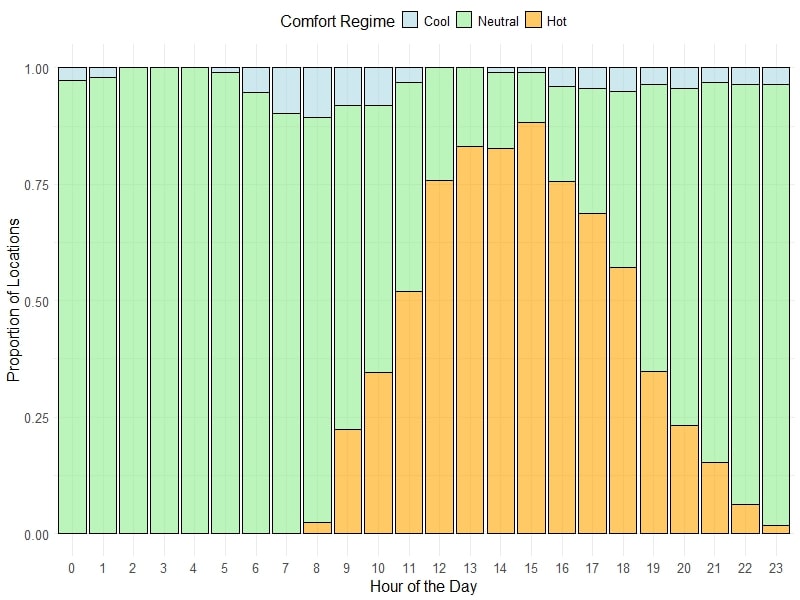}
    \caption{Hourly distribution of comfort regimes (Cool, Neutral, Hot) throughout the day. }
    \label{fig:barplot_state}
\end{figure}

Specifically, State 1 is characterized by lower average Temp, lower UTCI values, and higher levels of RF and RH. 
State 2 represents a more balanced environment with no RF, moderate Temp, and RH, along with lower WS. 
Notably, States 1 and 2 exhibit similar UTCI values, indicating that using UTCI alone as a thermal comfort index may not effectively differentiate between these phases.
State 3 is characterized by the highest Temp and UTCI values, alongside low RF and RH, making it the least thermally comfortable state. This pattern is reinforced by the distribution of the \textit{hour} feature, where 15:00 typically marks the hottest part of the day, followed by 03:00 and 08:00, reflecting the typical diurnal temperature cycle. As shown in Figure \ref{fig:barplot_state}, the hours around 15:00 have a greater number of stations clustered in the hot regime, further supporting the association between time of day and thermal discomfort.

The model spends 3.49\%, 65.89\%, and 30.62\% of its time in States 1, 2, and 3, respectively. This distribution indicates that the cool regime is relatively rare during the spring season in Singapore, with significant proportion of time spent in hot conditions. This frequency distribution reflects the tropical climate of Singapore.

\begin{table}[h]
\centering
\caption{State-conditional prototypes of features.}
\label{tab:state_conditional_prot}
\begin{tabular}{lccc}
\toprule
 & State 1 & State 2 & State 3 \\
\midrule
Temp (°C) & 25.51 & 28.19 & 30.51 \\
RH (\%) & 92.17 & 82.80 & 69.52 \\
RF (mm) & 0.43 & 0.00 & 0.01 \\
WS (m/s) & 4.55 & 3.06 & 4.94 \\
UTCI (°C) & 29.51 & 29.41 & 33.36 \\
Windy & 3 & 2 & 3 \\
Hour & 8:00 & 3:00 & 15:00 \\
\bottomrule
\end{tabular}
\end{table}

\begin{figure}[b]
    \centering
    \includegraphics[width=.8\linewidth]{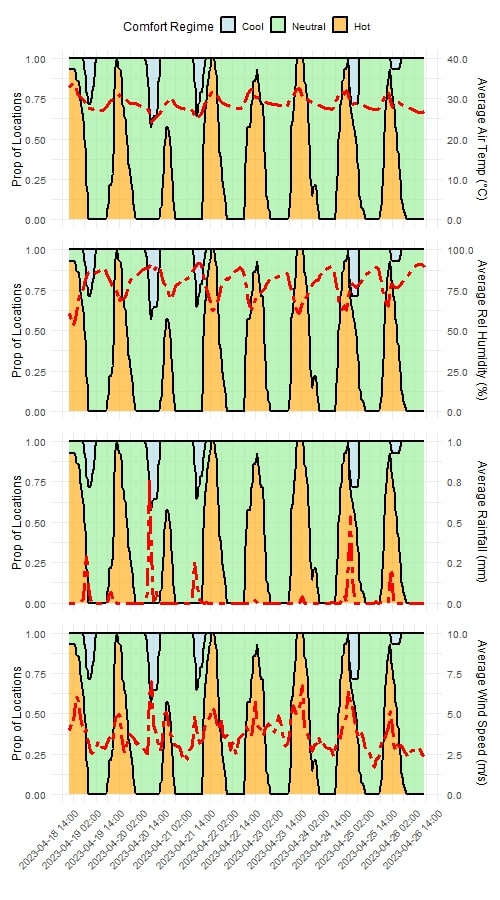}
    \caption{Temporal distribution of comfort regimes (Cool, Neutral, Hot) across locations. The red dashed line indicates the average air temperature, relative humidity, rainfall, and wind speed
    versus time.
    The left y-axis shows the proportion of locations classified into each comfort regime, while the right y-axis reflects the corresponding variables.}
    \label{fig:all_state}
\end{figure}

\begin{figure}[b]
    \centering
    \includegraphics[width=.8\linewidth]{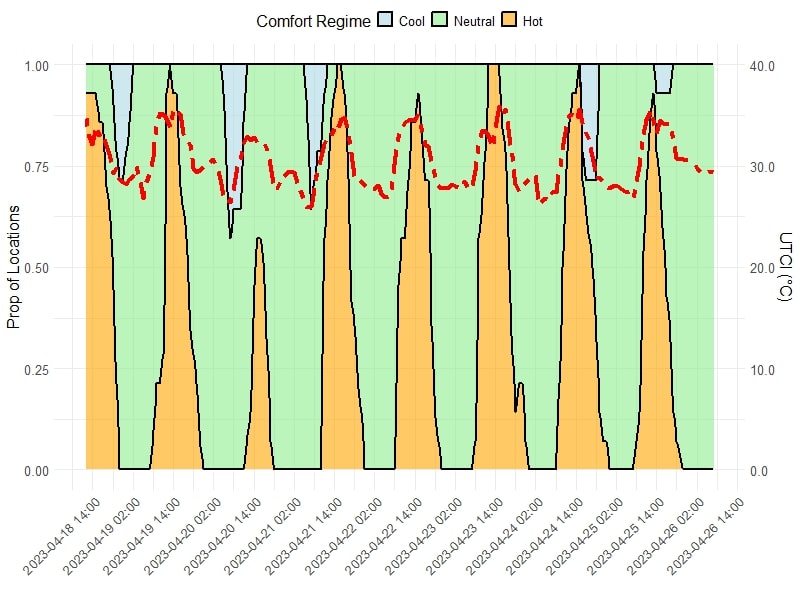}
    \caption{Temporal distribution of comfort regimes (Cool, Neutral, Hot) across locations, with the red dashed line representing the UTCI (°C) versus time. The left y-axis depicts the proportion of locations classified into each comfort regime, while the right y-axis indicates the UTCI.}
    \label{fig:utci_state_plot}
\end{figure}

Figure \ref{fig:all_state} shows the temporal distribution of comfort regimes (Cool, Neutral, Hot) across various locations, with the red dashed line depicting average values for Temp, RH, RF, and WS. Figure \ref{fig:utci_state_plot} presents a similar analysis but focuses on the UTCI.

As expected, Temp and UTCI show peaks corresponding to the hot regime, where higher Temp and UTCI values are closely associated with periods of increased thermal discomfort. The daily rise in RF after sunset is evident, and its role in influencing thermal comfort is clearly observed.

Figure \ref{fig:all_state} highlights the significant role of RF in outdoor thermal comfort, where peaks in RF are consistently followed by a higher probability of being in the cool regime. In fact,
RF helps lower air temperature by increasing moisture and reducing heat buildup \citep{morakinyo2019quantifying}.

Regarding WS, no clear periodic pattern emerges, except for a notable peak on April 20 at 5:00. This peak coincides with the start of a predominantly cool phase, highlighting a potential link between elevated WS and improved thermal comfort. 
However, WS does not consistently align with cool phases. This suggests that other factors like Temp and RH likely have a more significant influence on outdoor thermal comfort, as wind often feels stronger in  low-temperature climates, due to increased convective heat loss \citep{liu2023review}.

\begin{figure}[b]
    \centering
    \includegraphics[width=.8\linewidth]{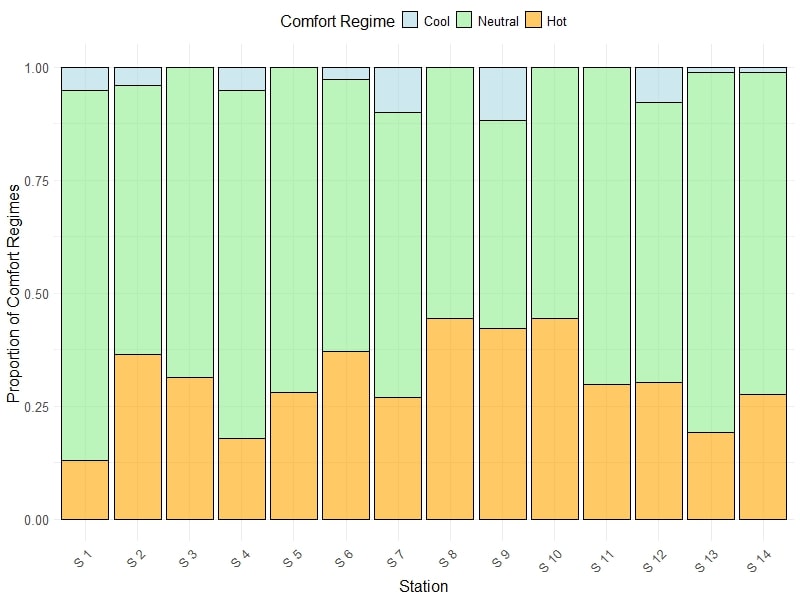}
    \caption{Proportion of comfort regimes (Cool, Neutral, Hot) across 14 weather stations.}
    \label{fig:stat_bar_plot}
\end{figure}

Figure \ref{fig:stat_bar_plot} shows the distribution of comfort regimes across the 14 weather stations (S1 to S14), highlighting spatial differences in the time spent in each comfort regime within the study area.
The results show that neutral conditions dominate across most stations, with certain stations, such as S8 and S10, exhibiting a higher proportion of time in the hot regime. In contrast, stations like S7 and S9 show a notable presence of the cool regime. Notably, Station S9 has the highest entropy, which is around 0.90.
This variation confirms that thermal comfort in urban areas is location-dependent, possibly influenced by local environmental factors like proximity to green areas, or urban heat island effects
\citep{hiemstra2017urban}. 

\begin{figure}[b]
    \centering
    \includegraphics[width=\linewidth]{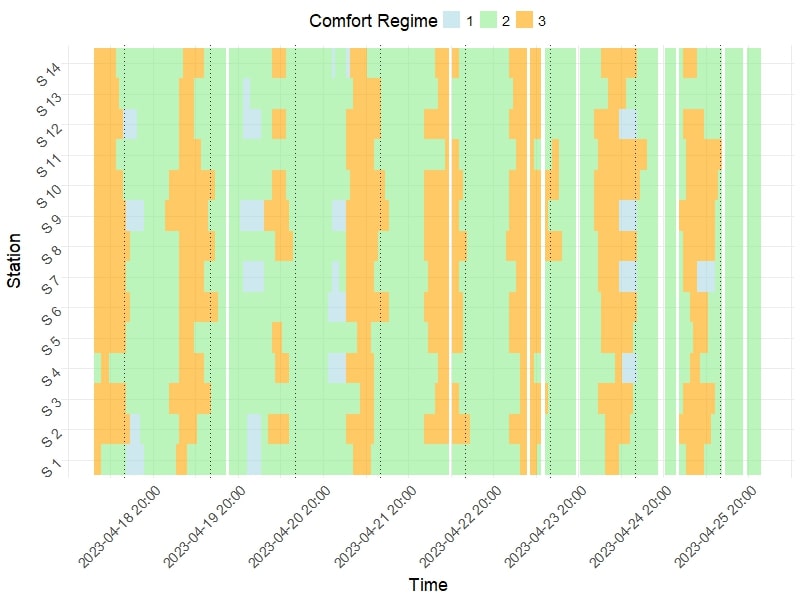}
    \caption{Heatmap of comfort regimes across the 14 weather stations.}
    \label{fig:heatmap_stat_time}
\end{figure}

Figure \ref{fig:heatmap_stat_time} illustrates the temporal distribution of comfort regimes across 14 weather stations over the study period. Each row represents a station, while the columns denote time intervals. The heatmap clearly highlights temporal variability, showing that certain stations tend to experience prolonged periods of thermal discomfort, particularly during certain hours of the day.
April 20th exhibits a dominant green color (neutral regime), which coincides with a spike in rainfall as previously discussed, highlighting the influence of precipitation on inferred comfort regime.
This representation helps in identifying both the temporal and spatial distribution of comfort regimes, and it can used as an instrument for monitoring urban microclimate dynamics.

\begin{figure}[b]
    \centering
    \includegraphics[width=.8\linewidth]{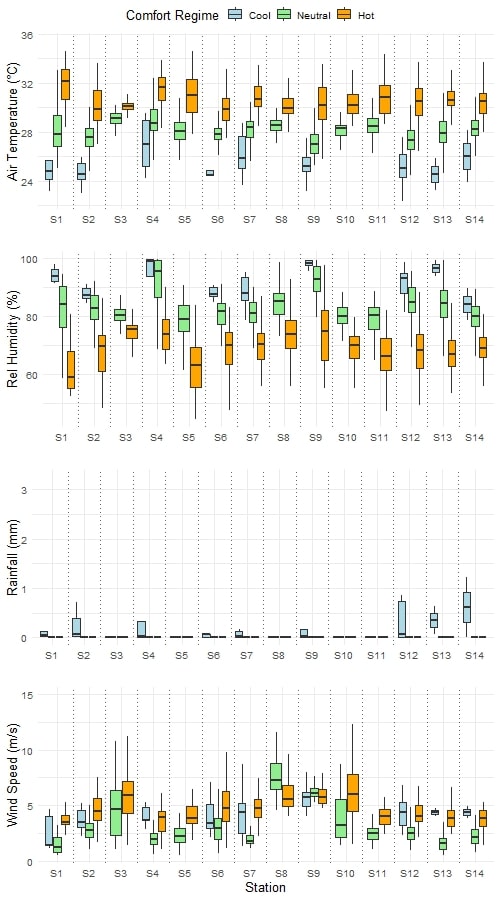}
    \caption{Boxplots of four weather variables—air temperature (°C), relative humidity (\%), rainfall (mm), and wind speed (m/s)—for each weather station (S1 to S14). The variables are grouped by comfort regimes, cool, neutral, and hot, indicated by light blue, light green, and orange colors, respectively.}
    \label{fig:boxplot}
\end{figure}

The boxplots in Figure \ref{fig:boxplot} show comparison of weather variables across the 14 stations for each comfort regime. S1, S4, and S5 exhibit higher median Temp in the hot regime, while S4 and S9 show consistently high RF, with significant variation across the different comfort regimes (cool, neutral, and hot). RF appears particularly prominent at S12 and S14, which are located outside the city center. These stations could be influenced by surrounding green spaces or lower urban density.
WS varies widely across the stations, with S8 and S10 recording the highest values across all regimes. However, no clear pattern emerges between WS and comfort regime, reinforcing that other factors such as temperature and humidity may play a larger role in defining thermal comfort in Singapore’s outdoor environments.

This spatial variability suggests the importance of considering local environmental factors, such as proximity to the coast or urban infrastructure, when assessing thermal comfort across urban areas.

\section{Discussion}
\label{sec:5}

In this work, we made two main contributions. First, we introduced a novel framework for spatio-temporal clustering that efficiently handles incomplete data, unequally sampled observations, and complex data structures. The framework is particularly suited for urban thermal comfort analysis, where environmental variables vary both spatially and temporally. Second, we applied this methodology to analyze outdoor thermal comfort in Singapore, using hourly weather data from 14 weather stations across the city.

Regarding the empirical application, we tested predictive ability comparing our clustering with thermal comfort feedbacks collected from users at different locations and times. This dataset consists of 54 responses to the question, ``Thermally, what do you prefer now?'' with options ``Warmer'', ``No change'', and ``Cooler''. 
Each feedback was matched with the median cluster from the closest weather station within a 4.92~km radius of the user's location at the time of feedback submission. This distance was selected as the minimum required to ensure that every feedback had at least one corresponding weather station cluster for comparison.
The model achieved an accuracy of 57\%, which is a reasonable outcome considering no physiological variables were used and no personal information on respondents, such as age, sex, and body mass index is available. 
In fact, thermal comfort is significantly influenced by physiological factors \citep{mamani2022variables}; however, our focus was on mapping the city's comfort zones based on environmental data alone.

We observed an increase in accuracy from 51\% to 57\% when the Universal Thermal Climate Index was included, emphasizing its importance. A key limitation is the lack of spatial resolution in the UTCI data, which could be addressed by using the SOLWEIG model \citep{lindberg2008solweig} to provide spatially detailed UTCI estimates and incorporate variables such as Sky View Factor (SVF) and Green View Factor (GVF).

Some limitations of the study include the small size of the user feedback dataset. We expect that more data would improve model accuracy. Regarding the methodology, further enhancements could include feature selection, potentially adapting the approach of \cite{nystrup:2021}, who propose a framework referred to as \emph{sparse statistical jump models}. 
Additionally, further research on hyperparameter selection is needed to refine the model.

\clearpage
\thispagestyle{empty}
\clearpage

\clearpage
\thispagestyle{empty}
\clearpage

\section*{Declaration}
\begin{itemize}
\item[] \textbf{Funding:} 
This work has been supported by the European Commission, NextGenerationEU, Mission 4 Component 2, ``Dalla ricerca all’impresa'', Innovation Ecosystem RAISE ``Robotics and AI for Socio-economic Empowerment'', ECS00000035.
\item[] \textbf{Conflict of interest/Competing interests:} The authors declare that there are no conflicts of interest.
\item[] \textbf{Data availability:} Data is available upon request from the corresponding author.
\end{itemize}

\bibliographystyle{Chicago}
\bibliography{biblio} 

\clearpage
\section*{Supplementary material}
This supporting information provides insights into the simulation study conducted to assess the clustering performance of our approach. Specifically, we present graphs illustrating the Balanced Accuracy (BAC) computed between true and estimated state sequences across all scenarios with varying values of the spatial penalty $\gamma$ and the temporal penalty $\lambda$.

\begin{figure}[b]
    \centering
    \includegraphics[width=.9\linewidth]{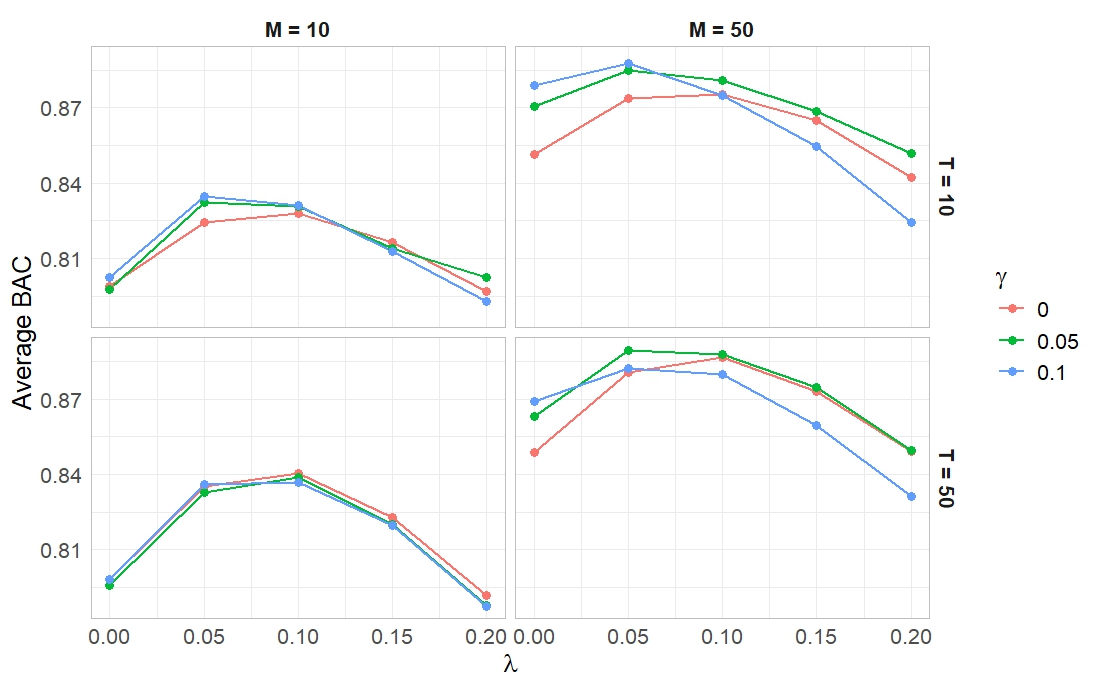}
 \includegraphics[width=.9\linewidth]{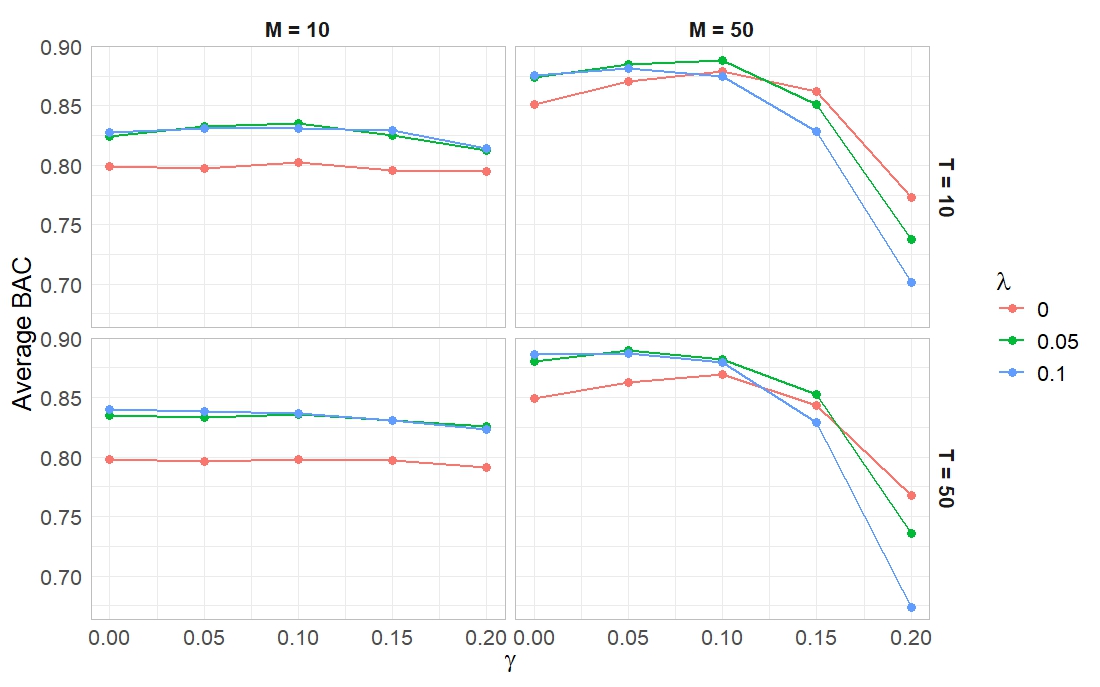}
    \caption{Average 
    Balanced Accuracy (BAC) computed between true and estimated latent sequences under 5\% missing data conditions, for \(P=10\).Each subplot presents BAC results across different values of $\lambda$ (top) and $\gamma$ (bottom), with each curve corresponding to a fixed level of $\gamma$ (or $\lambda$, respectively). }
    \label{fig:simstud_NA5_p10}
\end{figure}

\begin{figure}[b]
    \centering
    \includegraphics[width=.9\linewidth]{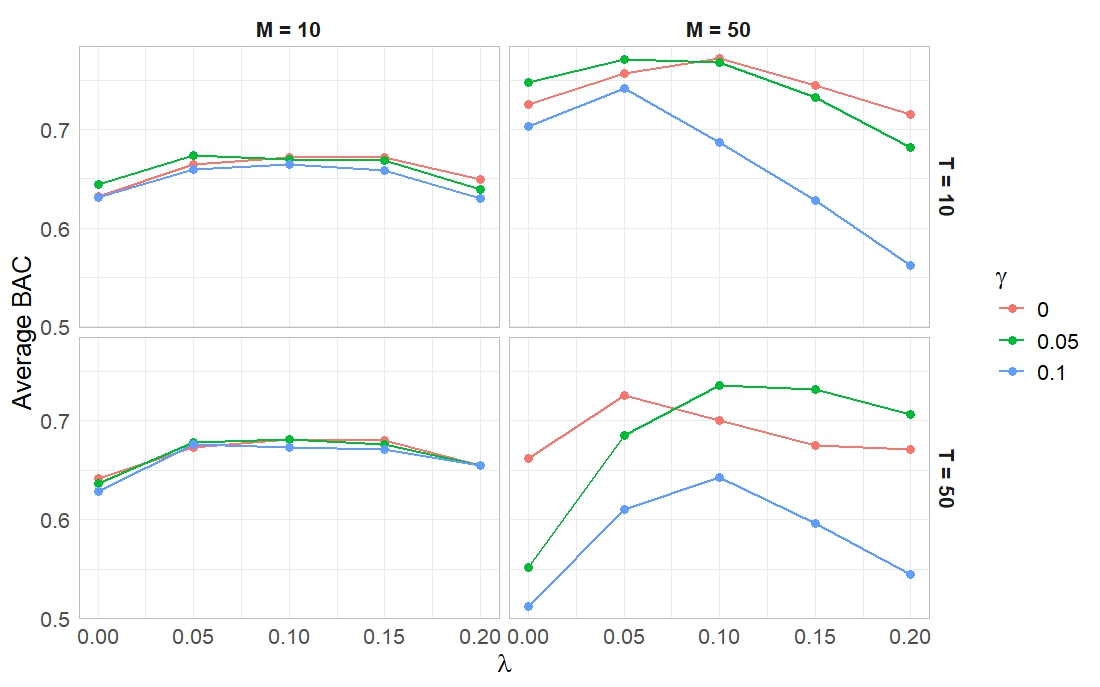}
 \includegraphics[width=.9\linewidth]{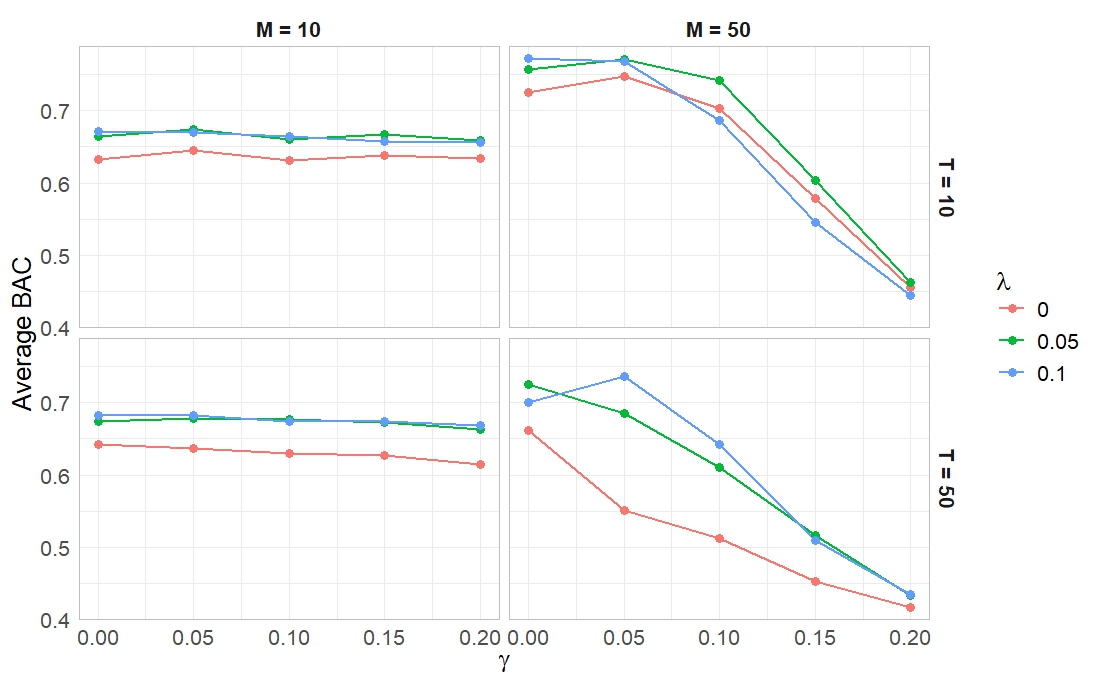}
    \caption{Average 
    Balanced Accuracy (BAC) computed between true and estimated latent sequences under 20\% missing data conditions, for \(P=10\).Each subplot presents BAC results across different values of $\lambda$ (top) and $\gamma$ (bottom), with each curve corresponding to a fixed level of $\gamma$ (or $\lambda$, respectively). }
    \label{fig:simstud_NA20_p10}
\end{figure}

Figures \ref{fig:simstud_NA5_p10} and \ref{fig:simstud_NA20_p10} present results for a feature count of $P=10$ with 5\% and 20\% missing data, respectively.

\begin{figure}[b]
    \centering
    \includegraphics[width=.9\linewidth]{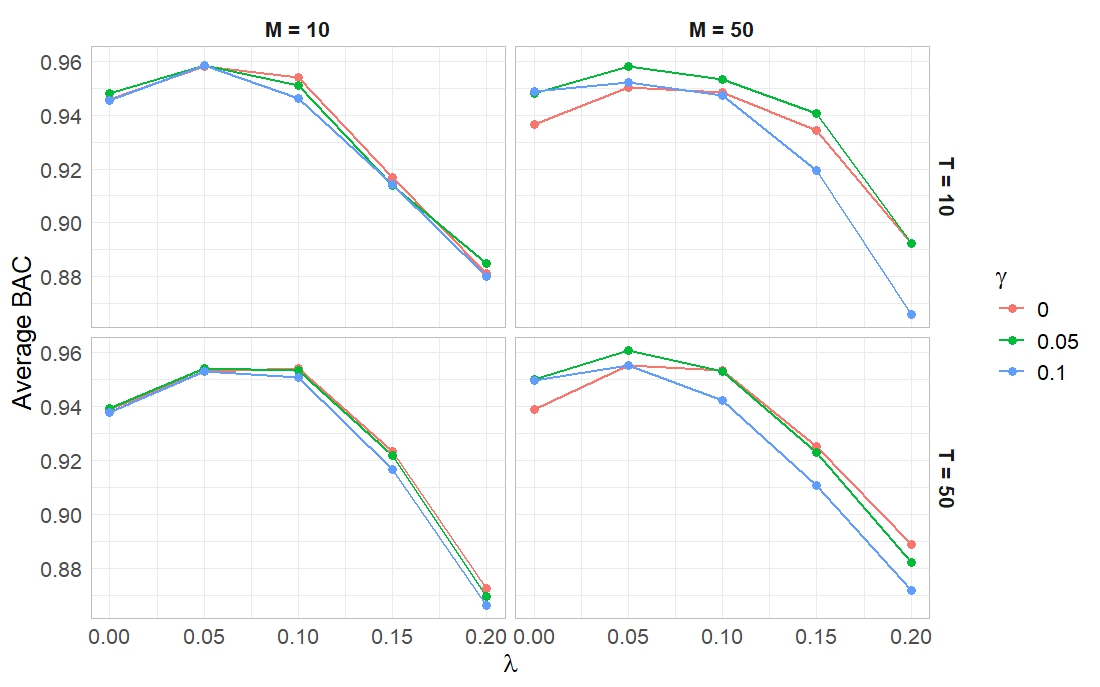}
 \includegraphics[width=.9\linewidth]{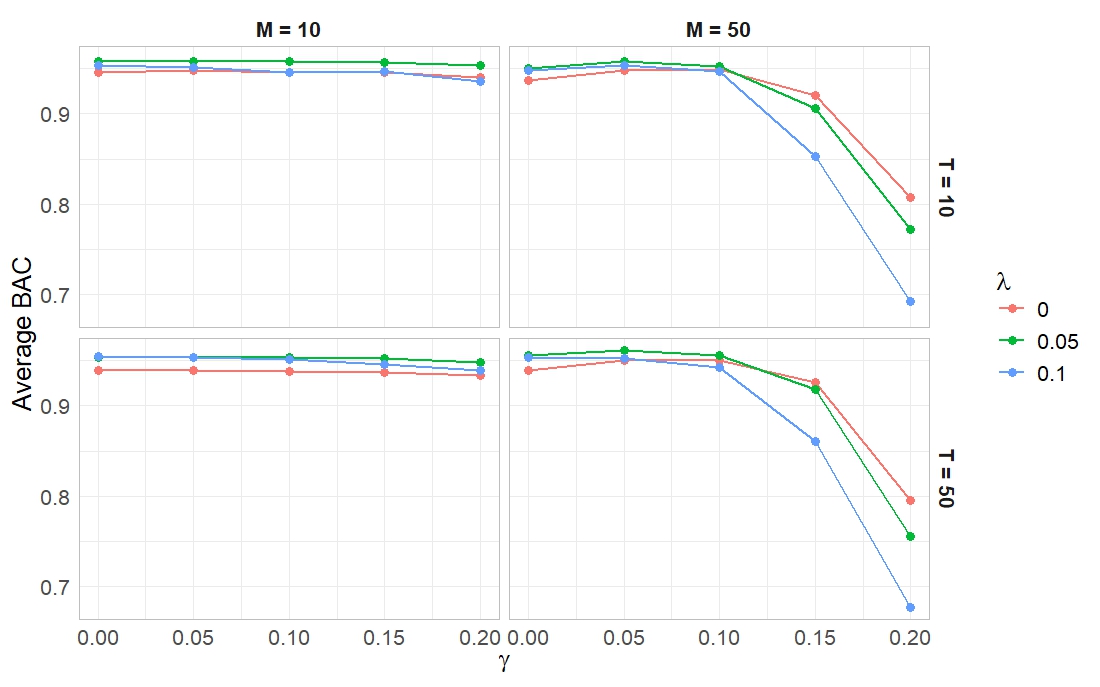}
    \caption{Average Balanced Accuracy (BAC) computed between true and estimated latent sequences, for \(P=20\), and for varying dimensions \( M \) and temporal lengths \( T \).Each subplot presents BAC results across different values of $\lambda$ (top) and $\gamma$ (bottom), with each curve corresponding to a fixed level of $\gamma$ (or $\lambda$, respectively). Missing data were introduced by randomly dropping 20\% of simulated time-series observations, preliminarily adjusting the total number of observations to
preserve dimensionality. }
    \label{fig:simstud_gaps_p20}
\end{figure}

Figure \ref{fig:simstud_gaps_p20} shows results when temporal gaps comprise 20\% of the total observations, with $P=20$.

\begin{figure}[b]
    \centering
    \includegraphics[width=.9\linewidth]{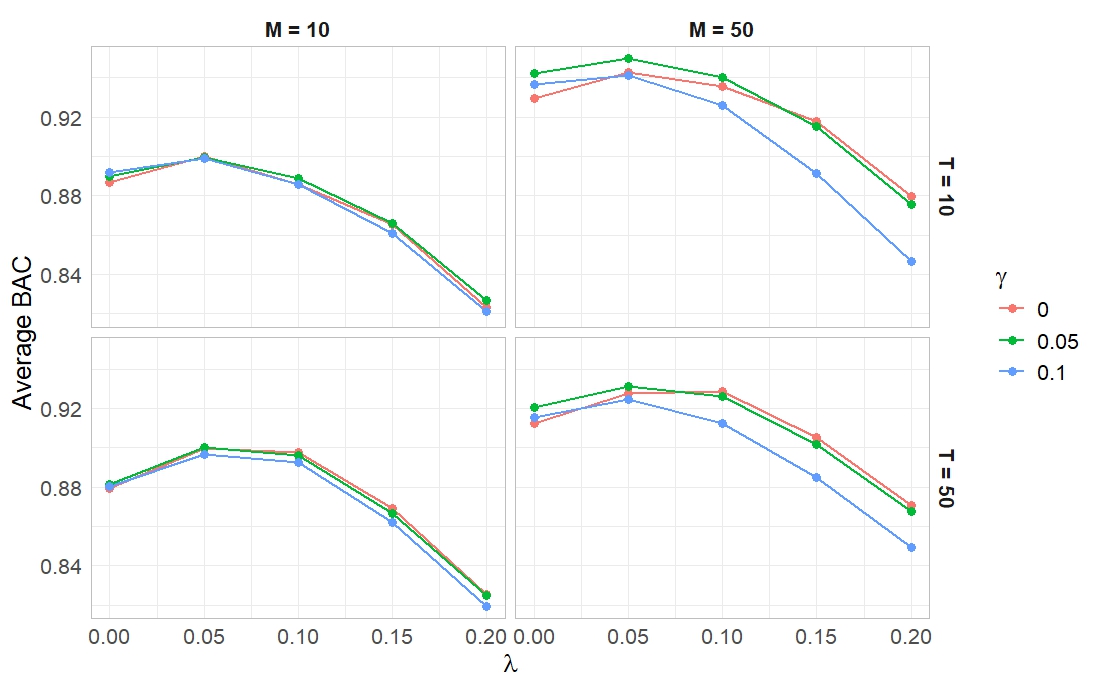}
 \includegraphics[width=.9\linewidth]{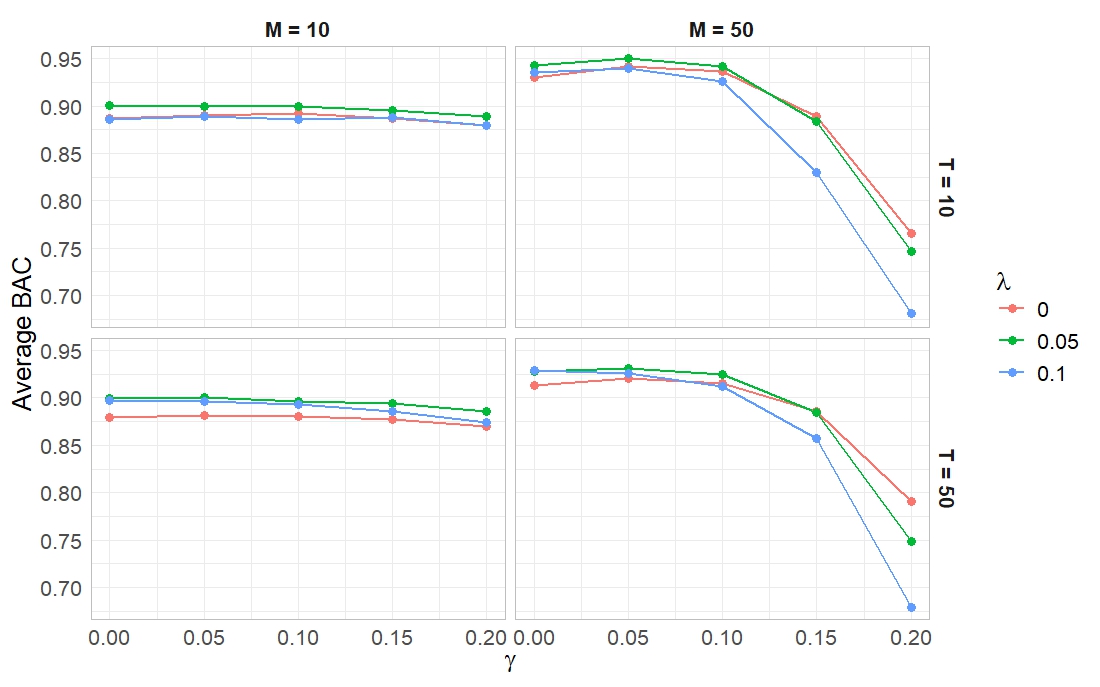}
    \caption{Average 
    Balanced Accuracy (BAC) computed between true and estimated latent sequences under 5\% missing data conditions, for \(P=20\).Each subplot presents BAC results across different values of $\lambda$ (top) and $\gamma$ (bottom), with each curve corresponding to a fixed level of $\gamma$ (or $\lambda$, respectively). }
    \label{fig:simstud_NA5_p20}
\end{figure}

\begin{figure}[b]
    \centering
    \includegraphics[width=.9\linewidth]{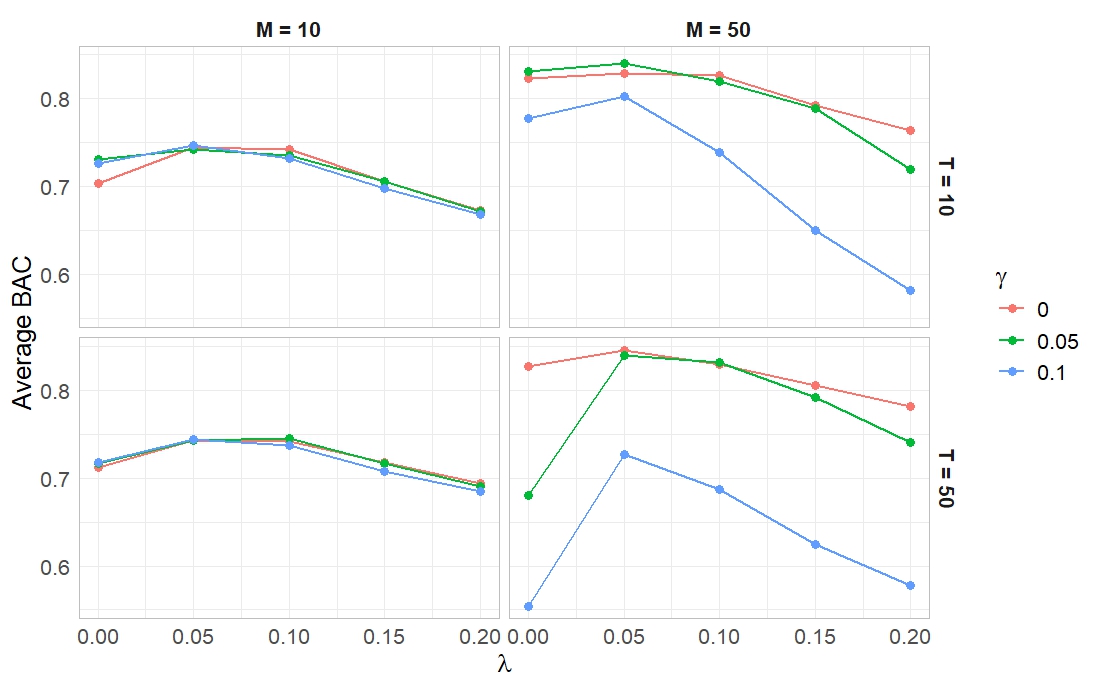}
 \includegraphics[width=.9\linewidth]{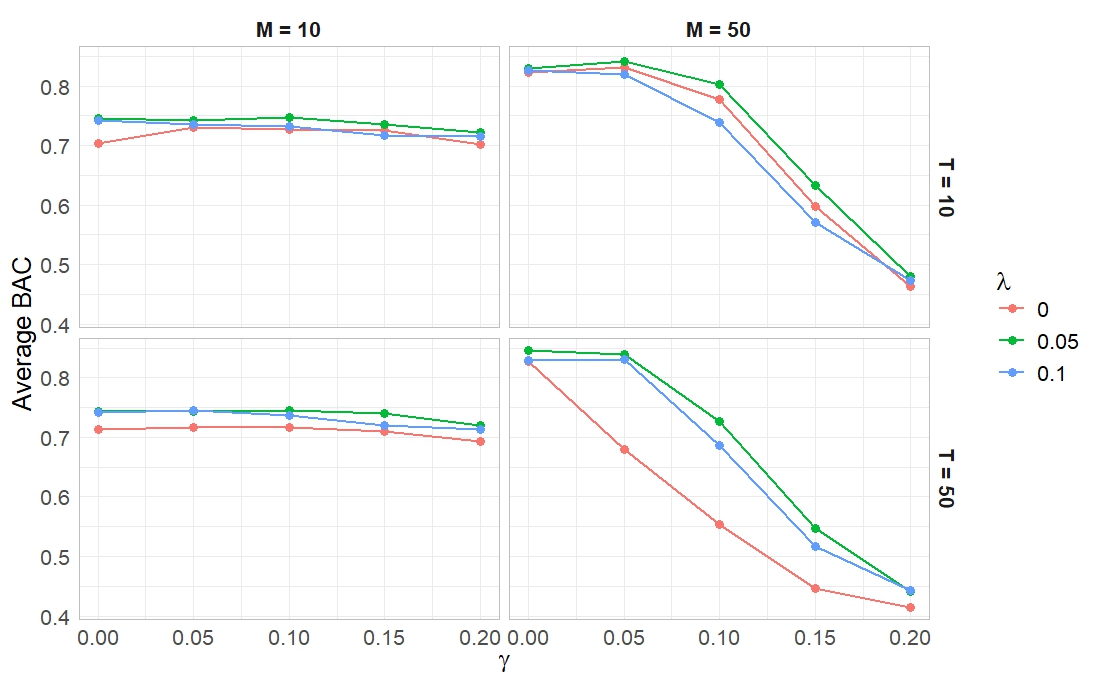}
    \caption{Average 
    Balanced Accuracy (BAC) computed between true and estimated latent sequences under 20\% missing data conditions, for \(P=20\).Each subplot presents BAC results across different values of $\lambda$ (top) and $\gamma$ (bottom), with each curve corresponding to a fixed level of $\gamma$ (or $\lambda$, respectively). }
    \label{fig:simstud_NA20_p20}
\end{figure}

Finally, Figures \ref{fig:simstud_NA5_p20} and \ref{fig:simstud_NA20_p20} display results for $P=20$ with 5\% and 20\% missing data, respectively.

\end{document}